# Environmental Lead-210 gamma-ray measurements in brain tumors

Michael S. Pravikoff (pravikof@cenbg.in2p3.fr), Christine Marquet (marquet@cenbg.in2p3.fr) and Philippe Hubert (hubertp@cenbg.in2p3.fr)
*Centre d'Études Nucléaires de Bordeaux-Gradignan (CNRS/Université de Bordeaux)*
*19 Chemin du Solarium*
*CS 10120*
*33175 Gradignan cedex, France*

ABSTRACT
In 2006 Momčilović et al. measured radioactivity levels of $^{210}$Po and $^{210}$Bi by $\alpha$- and $\beta$-measurements resp. to determine the natural distribution of environmental radon daughters in the different brain areas of an Alzheimer Disease victim. We show that $\gamma$-ray spectrometry of the mother nucleus $^{210}$Pb can be used. It is a direct measurement without tampering with the sample. However, because this technique is far less sensitive than $\alpha$- and $\beta$-spectrometry, activity values previously published were out-of-reach for $\gamma$-ray spectrometry. This is no longer the case with our ultra-low background high-purity Ge detectors which have been developed in the frame of the neutrino experiments. To ascertain this possibility, ten brain samples from Alzheimer patients and blanks as well as 10 samples of brain tumors have been investigated and results were cross-checked with two different spectrometers. Localization of the samples in the brain was not known to avoid biased interpretation of the results. Data taking lasted between 1 and 4 weeks for each sample. All ten Alzheimer/blank samples were below the limit of detection (LOD), 3 out of the 10 tumors lead to positive results (3.8 to 4.9 mBq/g of $^{210}$Pb) with a statistical significance of at least 3 standard deviations. Potential followup of this method is to enhance the detector sensitivity, having the spectrometer installed in an underground laboratory such as the one in Modane under the Alps where the background noise is 30-fold less than at the present location in Bordeaux. The final aim is to map a brain in collaboration with Alzheimer/Parkinson experts.

INTRODUCTION
In our environment, natural activity stems from the presence of several radioactive nuclei, among which the isotopes of Uranium ($^{238}$U, $^{235}$U), Thorium ($^{232}$Th) and their progeny, as well as Potassium ($^{40}$K). All these nuclei have decay half-lifes of the order of a billion years which are close to Earth's age. Our study focuses on the $^{238}$U decay chain (half-life = 4.5×10$^9$ years). Figure 1 shows the decay process from $^{238}$U all the way through $^{206}$Pb, a stable isotope.

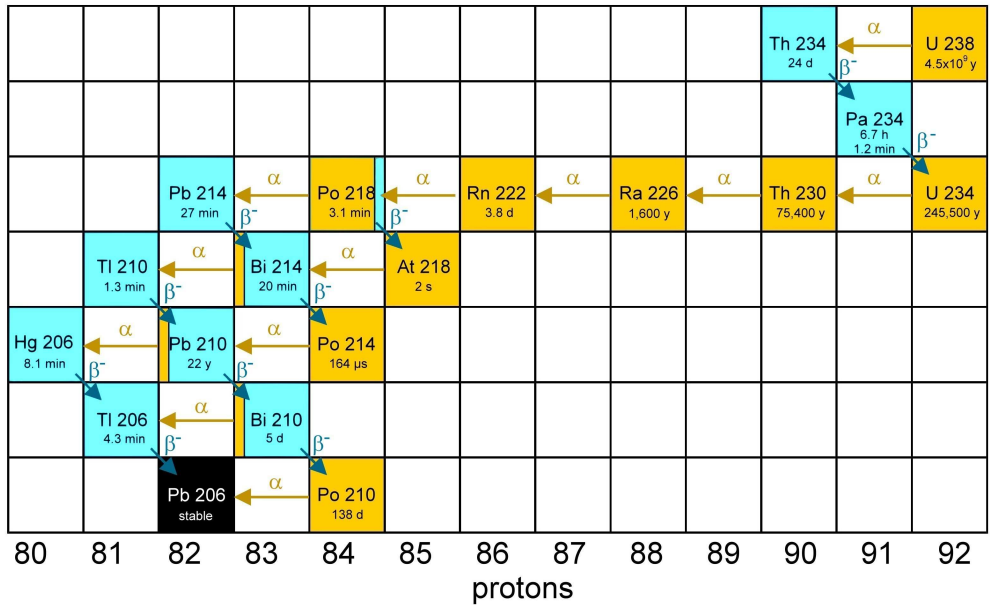

*Figure 1: $^{238}$U, one of the three natural radioactivity families*

$^{238}$U is found mainly in rocks and soils with a radioactive intensity around a few tens of Bq/kg. One of its progeny is $^{226}$Ra (Half-life = 1,600 years), a highly soluble element. Because of this property, it is found almost everywhere in nature at varying activity levels. This isotope is an alpha emitter and transforms to the well-known gaseous and radioactive Radon-222 ($^{222}$Rn) with a 3.8 day half-life. Being a neutral gas, it migrates from the rocks and soils to the atmosphere leading to Radon levels in the air ranging from a few Bq/m$^3$ to several thousands of Bq/m$^3$ in confined places like underground mining areas.

Radon-222 is also an alpha emitter and gives Polonium-218 ($^{218}$Po) in a positively charged state. Because it is ionized, it sticks on aerosols and/or dust particles present in the air. $^{218}$Po and progeny half-lifes are very short and the decay processes lead rapidly to Lead-210 ($^{210}$Pb) which has a half-life of 22 years, comparable to human life expectancy.

Earth gravity and rain precipitate $^{210}$Pb to the ground and over all that is growing. We have measured $^{210}$Pb activity in vine and oak tree leaves (around 100 Bq/kg dry weight), in fruit skins mainly from apples and peaches, in tobacco leaves (a few Bq/kg dry weight) and in seafood (high levels of $^{210}$Pb because of the presence of radium in water which is continuously filtered by the living organisms).

Lead-210 enters living organisms through breathing, eating and drinking. The amount is very low, however it is measurable by today's "low background level gamma spectrometers" like those from our research group, capable of measuring activities 3 to 4 magnitudes lower than the natural ambient level.

The human body rejects most of the ingested activity through natural processes, but some is left in the body, mostly in bones (around 2 Bq/kg), partly in soft tissues like liver, kidneys, lungs and muscles (around 0.2 Bq/kg). $^{210}$Pb is a quite harmless isotope: it is a beta emitter of low energy radiation ($Q_\beta$ = 63 keV), negligible compared to beta radiation from $^{40}$K and $^{14}$C, two isotopes naturally present in the human body at levels of some tens of Bq/kg. However $^{210}$Pb decays to Polonium-210 ($^{210}$Po), an alpha emitter with sufficient energy ($E_\alpha$ = 5.3 MeV) to damage DNA. Ingested at huge radioactivity levels of the order of 1 GBq, $^{210}$Po is lethal within a few weeks as in the famous case of the poisoning of Alexander V. Litvinenko. $^{210}$Po is even more radiotoxic than Plutonium: 1 µg [$^{210}$Po] = 0.16 Gbq = 14 kg [$^{238}$U].

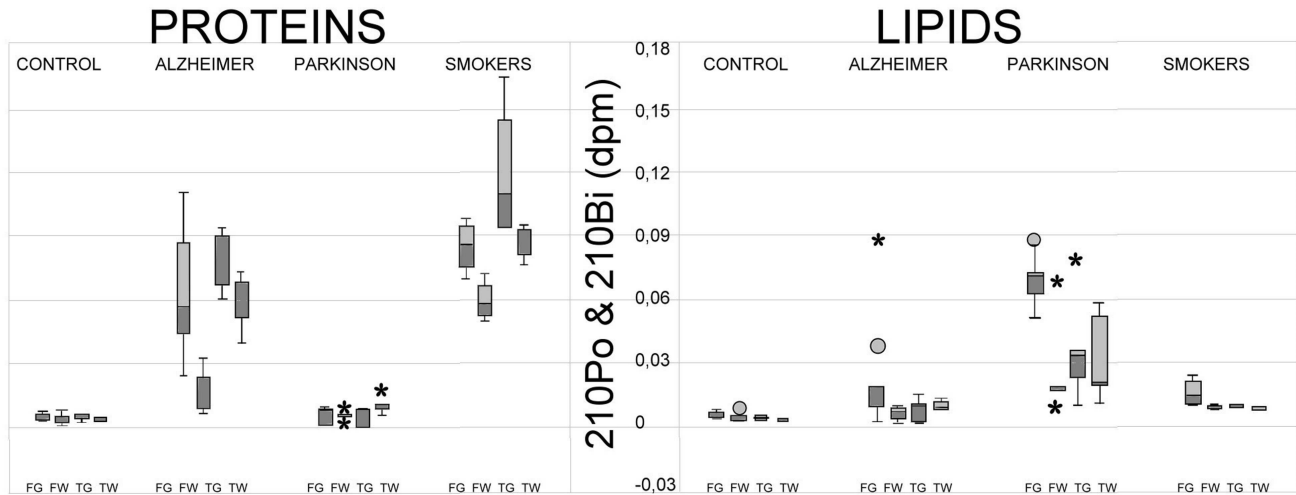

*Figure 2: radioactivity of 210Bi and 210Po inside the brain measured by Momčilović on a cohort of more than 30 patients suffering from Alzheimer or Parkinson diseases, smokers and non-smokers.*

*Radioactivity of $^{210}$Bi and $^{210}$Po inside the brain*
Numerous data compilations on the radioactivity levels in different parts of the human body are at hand. However, we are aware of only a single research team that has worked on the distribution of $^{210}$Po in the brain. In a 1999 published work, pursued in 2001, Momčilović et al. showed:

1/ the presence of radon progeny in human brains,

2/ that the activity levels were at least 10 times greater for brains from Alzheimer or Parkinson patients (even larger values were obtained for some smoker patients) than for control/blank brains,

3/ the accumulation of $^{210}$Bi/$^{210}$Po was not uniform in the brains, but seemingly dependent on the disease,

4/ the same distributions for $^{210}$Po (alpha measurement) and $^{210}$Bi (beta measurement) which implies the presence of their mother nucleus $^{210}$Pb.

Figure 2, which shows graphically those results, is reproduced from their paper.

In a second paper from 2006 following an oral communication in a conference in 2003, the same research team measured $^{210}$Po in different parts of a single old woman with Alzheimer Disease (AD), who passed away at 86 years of age. The observed activities are very weak, they range from 0.2 to

2 mBq for a 1 g sample, as shown in Figure 3, which comes from their paper.

Fortunately, $^{210}$Pb "regular" ingestion is only a few Bq on a daily basis. Nonetheless, and that is what triggered our concern, this is an on-going process all along one's life. What are the long-term effects on our health?

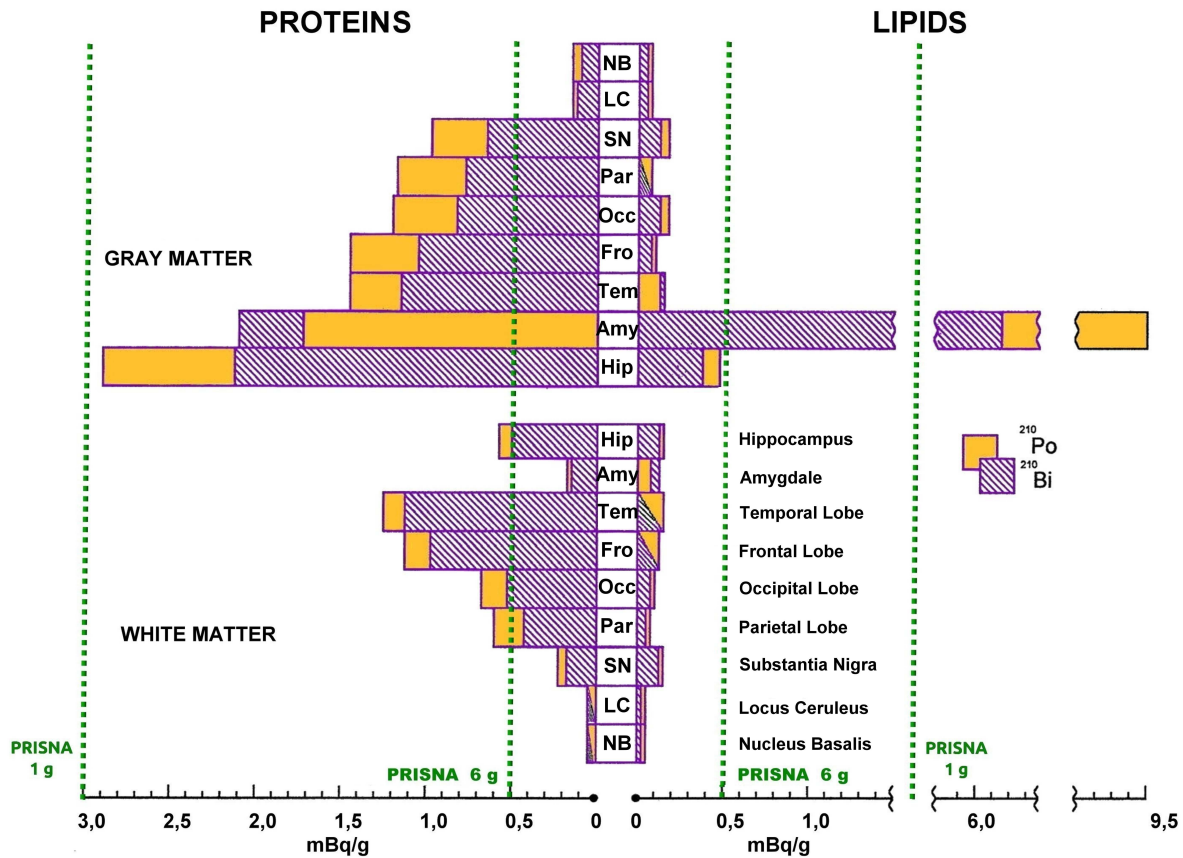

Figure 3: very fine analysis by Momčilović and al of the different parts of the brain of a deceased old woman, suffering from AD. Added to the original drawing are the LOD for our gamma-ray spectrometers at the PRISNA facility for bot a 1 g and a 6 g sample a a 1 week measurement.

EXPERIMENTAL
How to detect and measure $^{210}$Pb and its progeny ($^{210}$Bi and $^{210}$Po)?

There are several ways to reveal the presence of $^{210}$Pb in a given sample. The pioneering one relies on the chemical separation of $^{210}$Po followed by alpha spectrometry. This a very sensitive technique yielding results down to 0.1 mBq/g even with samples weighing less than a gram. The drawback is that the sample is destroyed and that the protocol demands a second measurement after a waiting period of 2 years to check if the secular equilibrium between mother and daughter nuclei was reached. On top of this, one needs radioactive tracers ($^{208}$Po or $^{209}$Po) which, nowadays, are difficult to obtain.

A second way is also a chemical separation, this time of ²¹⁰Bi, followed by liquid scintillation for beta spectrometry. The sensitivity is roughly the same as in the previous method, the protocol is complicated and the sample is destroyed. However, because of much shorter periods (half-lifes), the secular equilibrium is reached much quickly in this case (1-2 months instead of 2 years).

The third and last possibility: measure directly the 46 keV gamma-ray emitted by ²¹⁰Pb with a "low background noise gamma spectrometer". The advantages are three-fold: 1) no tampering with the sample which remains intact for further analysis and/or in case of a valuable specimen, 2) very light handling of the sample, 3) no need for a second measurement, data acquired is readily obtained and usable.

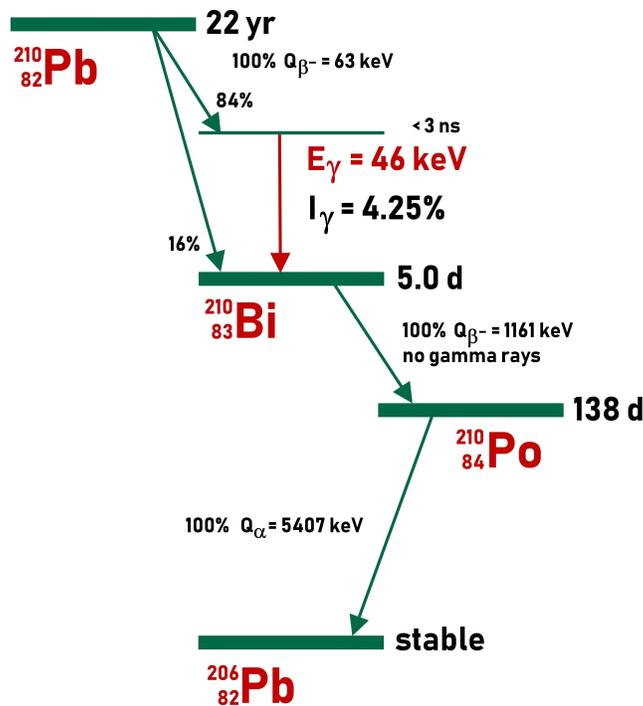

*Figure 4: ²¹⁰Pb decay scheme*

The only weak point compared to the two precedent techniques is that the measurement sensitivity is somewhat lower (around 1 mBq/g for a few grams of sample and one week data acquisition with our gamma spectrometers at the PRISNA[1] facility). This is due to the weak gamma-ray branching ratio for the 46 keV line (only 4.25% probability of decaying via the emission of a gamma ray). The partial decay scheme for ²¹⁰Pb is shown in Figure 4.

In Figure 3, we have added our experimental sensitivity (vertical green dotted lines) for a 300 cm³ Ge

---

1   PRISNA is a semi-underground facility, partially shielded from cosmic ray interactions and natural radioactivity, located on the premises of the CENBG institute in Gradignan

(Germanium) detector of "well type". This sensitivity depends on the sample mass and the duration of data acquisition: 3 mBq/g for a 1 g sample and 0.5 mBq/g for a 6 g sample. This means that we could have had positive results with the samples measured a decade earlier by Momčilović et al and this comforted us to get in touch with the La Pitié-Salpêtrière hospital in Paris, France for a first set of measurements and proof of feasibility.

RESULTS & DISCUSSION

Brain samples received weighed around 1 g and were conditioned either with or without formalin in sealed 5 cm$^3$ cylindrical polyethylene tubes. These tubes fit exactly the "well" part of our gamma-ray spectrometers.

Each sample was measured for 1 to more than 3 weeks to minimize statistical uncertainties. Since this was a proof-of-feasibility experiment, we cross-checked the data by performing dual measurements on a second detector in order to ascertain that the observed radiation was not inherent to a specific spectrometer. For the 46 keV line, the detector's efficiency is 60% with an energy resolution (FWHM/Full Width Half Maximum) of 2 keV.

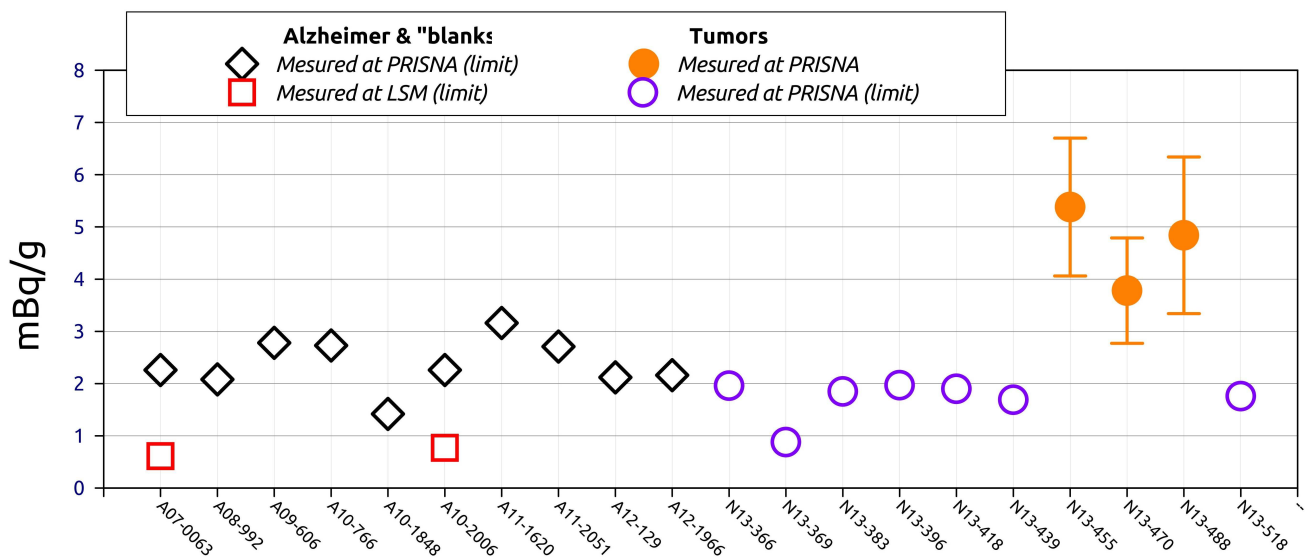

*Figure 5: graphical representation of all measurements from Table 1. Limit of detection (LOD) is around 3 mBq/g for a 1 g sample and a one week data acquisition.*

We did not have information on the samples that were sent. Especially, no clue was given as to which part of the brain they stemmed from.

Two series were obtained. The first one, marked as A07-*xxx* to A12-*xxx* included 5 "blank" and 5 "Alzheimer" samples. They were conditioned with formalin in the polyethylene tubes. The second series, noted N13-*xxx*, originated from patients with different types of brain tumors. They were conditioned without formalin. The samples weighed between 0.77 g and 1.84 g.

Table 1 lists the samples and the measurements data. Figure 5 is a graphical representation of the results. All the A*xx.xxx* samples lead to a detection limit of around 2 to 3 mBq/g. Two of those samples, A007-0063 and A10-2006, were also measured at the LSM (Laboratoire Souterrain de Modane under the Alps), but only for one week each. The results were also limit values, resp. < 0.61 and < 0.77 mBq/g with a sensitivity threshold 5 times better than in PRISNA, thanks to better shielding due to thousands of feet of rock which lowers the background by a factor 30 compared to PRISNA.

With the N13-*xxx* series, 7 samples were also at the detection limit (values between 1.7 and 3.4 mBq/g, depending on sample mass and measurement duration). Most interesting were the measured positive values for 3 samples, between 3.8 and 4.9 mBq/g, with a statistical significance of at least 3 standard deviations.

From Figure 5 one infers that the detection limit in PRISNA is around 3 mBq/g for a 1 g sample and a one week data acquisition.

CONCLUSIONS

This preliminary study confirmed that today's sensitivity of the gamma spectrometers in PRISNA is good enough to measure $^{210}$Pb in brain samples originating from the natural radioactivity. At this stage of the study, lack of knowledge of the precise localization in the brain of the samples prevents us from associating our results with a pathology.

Upcoming improvements are:

1/ enhance the detector sensitivity with a new "well type" Ge gamma-ray spectrometer of a smaller volume (100 cm$^3$ instead of 300 cm$^3$), which decreases the interfering high energy gamma ray contribution (Compton effect) and has excellent resolution (1 keV at 46 keV) with the same efficiency (60% at 46 keV).

2/ have the spectrometer installed at LSM in the underground laboratory with a 30-fold decrease in background noise.

Last but not least, setting up and broadening a collaboration with Alzheimer/Parkinson specialists.

ACKNOWLEDGMENTS

The authors wish to acknowledge the precious help they received from neuro-specialists, especially from the Neuro-CEB from the La Pitié-Salpêtrière hospital in Paris, France, which provided us with the samples.

One of us, Ph. Hubert, is redeemable for many interesting talks with Dr G.I. Lykken, who pioneered this work.

We are also grateful to the IRSN which allotted us a couple of weeks for measurements with their gamma-ray spectrometer at the LSM.

BIBLIOGRAPHIC REFERENCES

- Momčilović B. et al., 2001, *Environmental Lead-210 and Bismuth-210 accrue selectively in the brain proteins in Alzheimer disease and brain lipids in Parkinson disease*, Alzheimer Diseases and Associated Disorders, 15 (2), pp. 106-115
- Momčilović B. et al., 1999, *Environmental radon daughters pathognomonic changes in the brain proteins and lipids on patients with Alzheimer's disease and Parkinson's disease, and cigarette smokers*, Arh. hig. rada. toksikol., 50 (4), pp. 347–369
- B. Momčilović and G.I. Lykken, 2003, *Distribution of $^{210}$Po and $^{210}$Bi radon daughters in the brain*

*proteins of a subject who suffered from Alzheimer's disease*, V. simpozij HDZZ, Stubičke Toplice, 2003
- B. Momčilović, G.I. Lykken and M. Cooley, 2006, *Natural distribution of environmental radon daughters in the different brain areas of an Alzheimer Disease victim*, Molecular Neurodegeneration 2006, 1, pp.11-17

Lexicon
| | |
|---|---|
| CENBG | Centre d'Études Nucléaires de Bordeaux-Gradignan |
| IRSN | Institut de Radioprotection et de Sûreté Nucléaire |
| LSM | Laboratoire Souterrain de Modane |
| Neuro-CEB | Neuro-Cérébrothèque (a tissue bank for neurological research) |
| PRISNA | Plate-forme Régionale Interdisciplinaire de Spectrométrie Nucléaire en Aquitaine |
| AD | Alzheimer Disease |

| | spectrometer location | sample ID | measurement duration (days) | weight (g) | activity (mBq/g) |
|---|---|---|---|---|---|
| **ALZHEIMZER** | PRISNA | A07-0063 | 10.9 | 1.14 | 2.3* |
| | LSM | | 6.5 | | 0.6* |
| | PRISNA | A08-992 | 10.3 | 1.38 | 2.1* |
| | | A09-606 | 8.0 | 1.00 | 2.8* |
| | | A10-766 | 7.6 | 0.99 | 2.7* |
| | | A10-1848 | 8.9 | 1.84 | 1.4* |
| | PRISNA | A10-2006 | 12.2 | 0.91 | 2.3* |
| | LSM | | 6.5 | | 0.8* |
| | PRISNA | A11-1620 | 8.0 | 1.00 | 3.2* |
| | | A11-2051 | 7.1 | 1.09 | 2.7* |
| | | A12-129 | 7.9 | 1.30 | 2.1* |
| | | A12-1966 | 8.4 | 1.23 | 2.2* |
| **TUMORS** | PRISNA | N13-366 | 20.7 | 0.77 | 2.0* |
| | | N13-369 | 33.2 | 1.50 | 0.6* |
| | | N13-383 | 7.0 | 1.63 | 1.8* |
| | | N13-396 | 13.9 | 1.08 | 2.0* |
| | | N13-418 | 21.0 | 1.45 | 1.9* |
| | | N13-439 | 9.0 | 1.34 | 1.7* |
| | | N13-455 | 14.0 | 0.83 | 4.9 (1.3) |
| | | N13-470 | 21.0 | 0.95 | 3.8 (1.0) |
| | | N13-488 | 22.1 | 0.90 | 4.8 (1.5) |
| | | N13-518 | 10.9 | 0.92 | 1.8* |

*Table 1: list of samples, measurement duration (days), sample weight, values obtained or limits (marked by \*) for $^{210}$Pb (mBq/g). All measurements done at the PRISNA facility in Gradignan (Bordeaux*

*suburb). Two samples were measured at both PRISNA and the LSM (Laboratoire Souterrain de Modane), a laboratory in a road tunnel between Italy and France, under the Alps.*